# Light tailored by multimode fiber for multiphoton fluorescence microscopy

**Raphaël Jauberteau[1,*], Alessandro Tonello[2], Yifan Sun[1], Mario Zitelli[1], Mario Ferraro[1], Fabio Mangini[1], Pedro Parra-Rivas[1], Tigran Mansuryan[2], Yago Arosa[2,3], Vincent Couderc[2], Stefan Wabnitz[1,4]**

[1]*Department of Information Engineering, Elect. and Telecommunications, Sapienza University of Rome, Via Eudossiana 18, 00184 Rome, Italy*
[2]*XLIM, UMR CNRS 7252, Université de Limoges, 123 Avenue A. Thomas, 87060 Limoges, France*
[3]*Departamento de Física Aplicada, Universidade de Santiago de Compostela, Praza do Obradoiro 0, 15705 Santiago de Compostela, Spain*
[4]*CNR-INO, Istituto Nazionale di Ottica, Via Campi Flegrei 34, 80078 Pozzuoli, Italy*
**raphael.jauberteau@gmail.com*

**Abstract:** We study the diffraction of a particular class of beams, composed only by a combination of azimuthally invariant guided modes of an optical fiber. We demonstrate that such beams can be obtained by injecting a Gaussian beam in a small piece of silica graded-index multimode fiber. This minimalistic low-cost method is applied for improving the axial resolution of a two-photon microscope.

## 1. Introduction

Multimode fibers (MMFs) have held the attention of the scientific community over the past few years, because of their involvement in both fundamental physics research [1-3] and several new technological applications, such as white light source generation [4], micro-machining [5], new fiber laser sources [1], space division multiplexing for optical communications [6], and high-resolution biomedical imaging [7,8]. Specifically, by using MMFs, it has been shown that it is possible to shape and deliver a laser beam for microscopy and endoscopy experiments. In addition, most of these imaging techniques require a careful control of the beam before its injection into MMFs, and rely on a spatial light modulator to couple the initial pump light, which significantly slows the rate of image acquisition [9-11].

However, several years ago it has been shown that it is possible to modify the shape of a beam by using nonlinearity in graded index (GRIN) MMFs, without the need of any physical constraints or phase shaping before its injection in the fiber. This effect, called beam self-cleaning, appears above a certain power threshold, and leads to modal energy exchanges in the GRIN MMFs, which reorganize the central portion of the beam into a bell shape [12], while a small part of the light remains coupled in high-order modes, depending on the injection conditions [13]. This particular effect, which is enabled by a natural self-imaging process, was exploited by Moussa et al. to reach a high spatial resolution in an imaging system based on standard 50/125 μm MMF [7]. They also demonstrated the possibility to build a microscope based on this ultrafast beam self-shaping, which preserves a high image acquisition rate. This is due to the use of the Kerr response of the fiber, without resorting to any electro-optical spatial shaping devices. In the same reference [7] authors highlight the robustness of the output beam profile versus fiber manipulations, which improves image stability for two- and three-photon fluorescence imaging. However, Krupa et al. [2] underlined that a centered and axial Gaussian excitation of a GRIN MMF may generate, in the linear regime of propagation, a multimode beam composed of symmetric low-order modes with a dominant Gaussian near-field profile but with a beam divergence larger (Rayleigh length shorter) than that of a pure Gaussian beam.

Multiphoton microscopy is a type of fluorescence imaging based on nonlinear absorption [14,15] for probing a sample with a near-infrared (NIR) laser as a light source. Multiphoton microscopy possesses many advantages, when compared with classic linear fluorescence microscopy. It improves depth penetration, since the scattering of NIR light is lower than that of visible light in biological tissues. The multiphoton absorption probability is also very low, meaning that such a process can only appear at the focal plane of the objective (where the light intensity is maximum), thus improving the spatial transverse resolution of the microscope, and highly diminishing photo-bleaching and photo-toxicity. Fluorescence imaging with customized beams has been extensively studied, using control of amplitude, phase, polarization... [16], improving performance in terms of spatial resolution, signal-to-noise ratio, contrast and imaging speed. However, there is no magic beam with a single shape. Each approach has advantages and disadvantages, which need to be properly balanced for a given application.

In this work, we show that the free space diffraction of a multimode beam, constituted of symmetric transverse modes excited in GRIN MMF, can be used for improving the axial resolution of a two-photon fluorescence microscope. Due to the relatively large core diameter of GRIN MMF, when an initial Gaussian beam is launched, it tends to generate not only the fundamental mode but also higher-order modes. This occurrence results in a linear combination of modes, leading to a periodic evolution of the optical beam as it travels along the MMF. It is important to note that this periodic beam evolution is distinct from the intensity-dependent output beam reshaping described in Ref. [7]. Our work is a novelty since radial modes are often associated with Bessel or Bessel-type beams [17], which are generated through other methods (most of the time using an axicon) and which possess a longer Rayleigh length. They have other applications in microscopy, where collimation over a long distance is required. For example, Bessel beams are used to enhance light sheet microscopy [18], where a beam with very low divergence is required along one of the transverse axes. These Bessel beams have a given spatial phase distribution, ensuring coherent extension in close vicinity to the waist position.

In our experiment, because of on-axis excitation, the beam is mainly composed of a coherent sum of symmetric modes of the GRIN MMF. The phase difference among these modes leads to complex spatial interferences. As a result, the Rayleigh range of the output beam is then reduced, in comparison to a Gaussian beam. Owing to the graded profile of the fiber core refractive index, the output beam can be refocused outside the fiber by a simple spherical converging lens. Thus, the parabolic phase introduced by the fiber on the propagating multimode beam can be counterbalanced. Similar beamforms were obtained by Zhu et al. by splicing a single-mode fiber with a multimode one [19]. However, splicing two different types of fibers may lead to back reflections, that could damage the input laser. Here we show that it is possible to generate these beams by using a single piece of multimode GRIN fiber. Thus, real-time repositioning of the incident beam can be realized, in order to improve the spatial longitudinal resolution in a multiphoton imaging experiment. This simple and low-cost method allows for the generation of a multimode beam composed of the coherent sum of the guided radial modes which, when injected in the multiphoton microscope, gives a twofold improvement of the longitudinal resolution; at the focal plane of the microscope objective, the beam shape of the radial multimode beam appears as a Gaussian, but with a shorter Fresnel length.

## 2. Simulation

To prove the potential of such a specifically shaped beam in two-photon fluorescence microscopy, we simulated first the free-space propagation of a beam composed of the sum of the first 16 Laguerre-Gaussian $LG_{0x}$ modes only. The first 9 modes ($LG_{01-09}$) that compose this multimode beam are visible in Fig. 1-square insets, where the mode weight in amplitude are respectively; $LG_{01}$ : 1, $LG_{02}$ : 10, $LG_{03}$ : 100, $LG_{06}$ and $LG_{016}$ : 10000, and the rest of the $LG_{0x}$ 's : 1000. This multimode beam provides a good approximation of the output of the GRIN fiber.

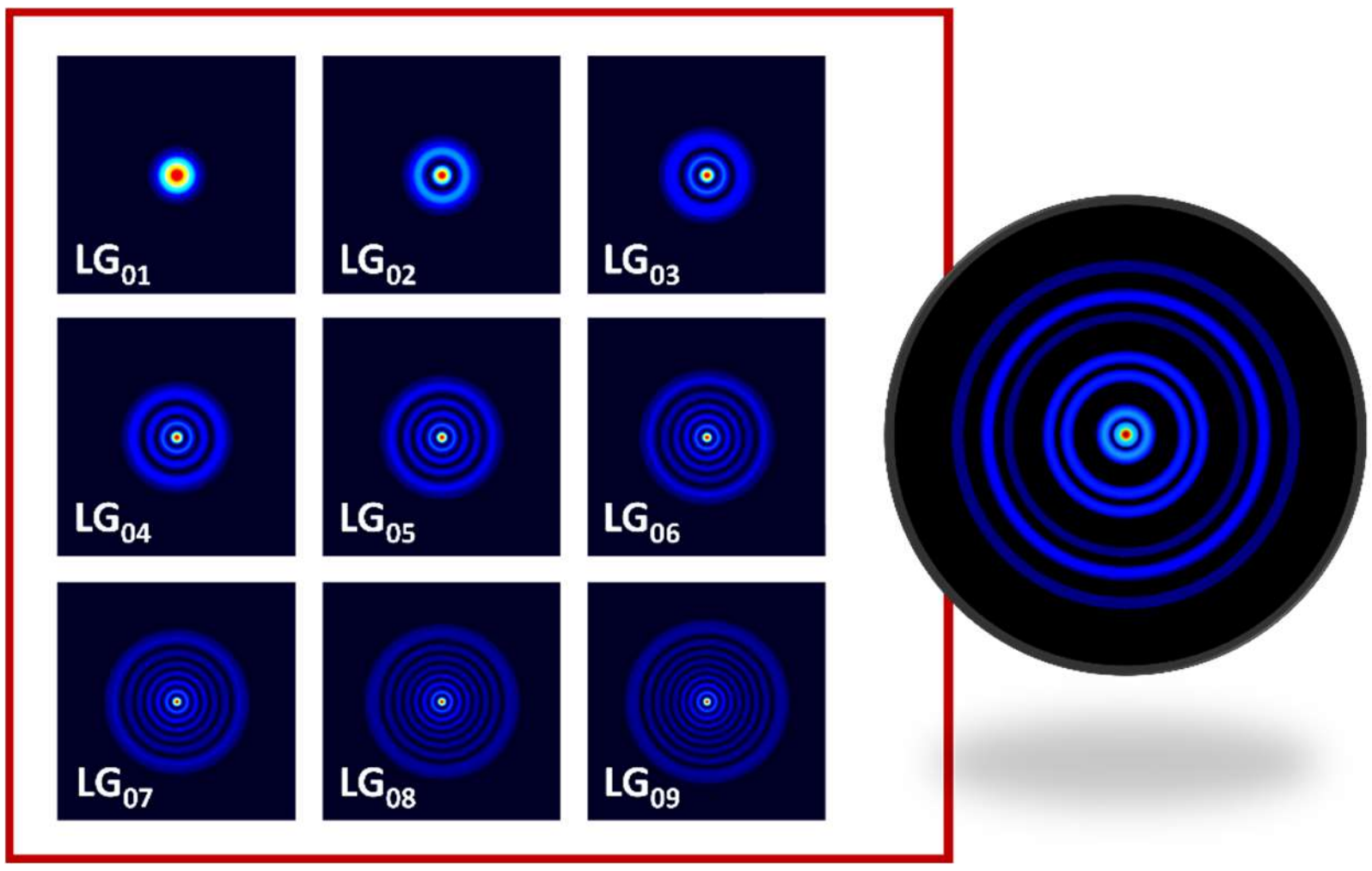

Fig. 1. Modal composition (square insets) of the simulated radial multimode beam intensity (round inset).

Its evolution in free space is shown in Fig. 2(b), where we also compare its propagation with that of a pure Gaussian beam with the same diameter, when measured at full-width-half maximum in intensity (FWHMI) (fig 2(a)). These simulations are obtained with a code based on the finite-element beam propagation method [20]. To match the experimental setup, the beams have a wavelength of 1030 nm. We applied these results to two-photon microscopy, by calculating the propagation of the squared value of the simulated beam intensity, and then filtering out the low-intensity values that would not generate a two-photon fluorescence signal (as visible on the round inset – fig 2(b)). From the simulation, we observed a decrease in the Rayleigh length of the multimode beam (Fig. 2 (b)), when compared with that of the Gaussian beam (Fig. 2 (a)), and this for the same beam diameter (FWHMI) at $z = 0$ mm. For the sake of comparison between the two beams propagation, we consider an arbitrary threshold, that we set at 30 % of the maximum of light intensity. Whereas the corresponding two-photon fluorescence detection threshold will depend on many parameters, such as the laser pulse energy or the detection device efficiency. We observe from Fig. 2(b) that the expected range of two-photon excitation of the multimode beam is almost halved with respect to that of a Gaussian beam. By using such multimode beam rather than a Gaussian beam, one can expect similar improvement in the microscope axial resolution. The improvement factor will depend on the modal composition of the multimode beam: the higher the number of modes, the shorter the Rayleigh length and, consequently, the greater the axial resolution.

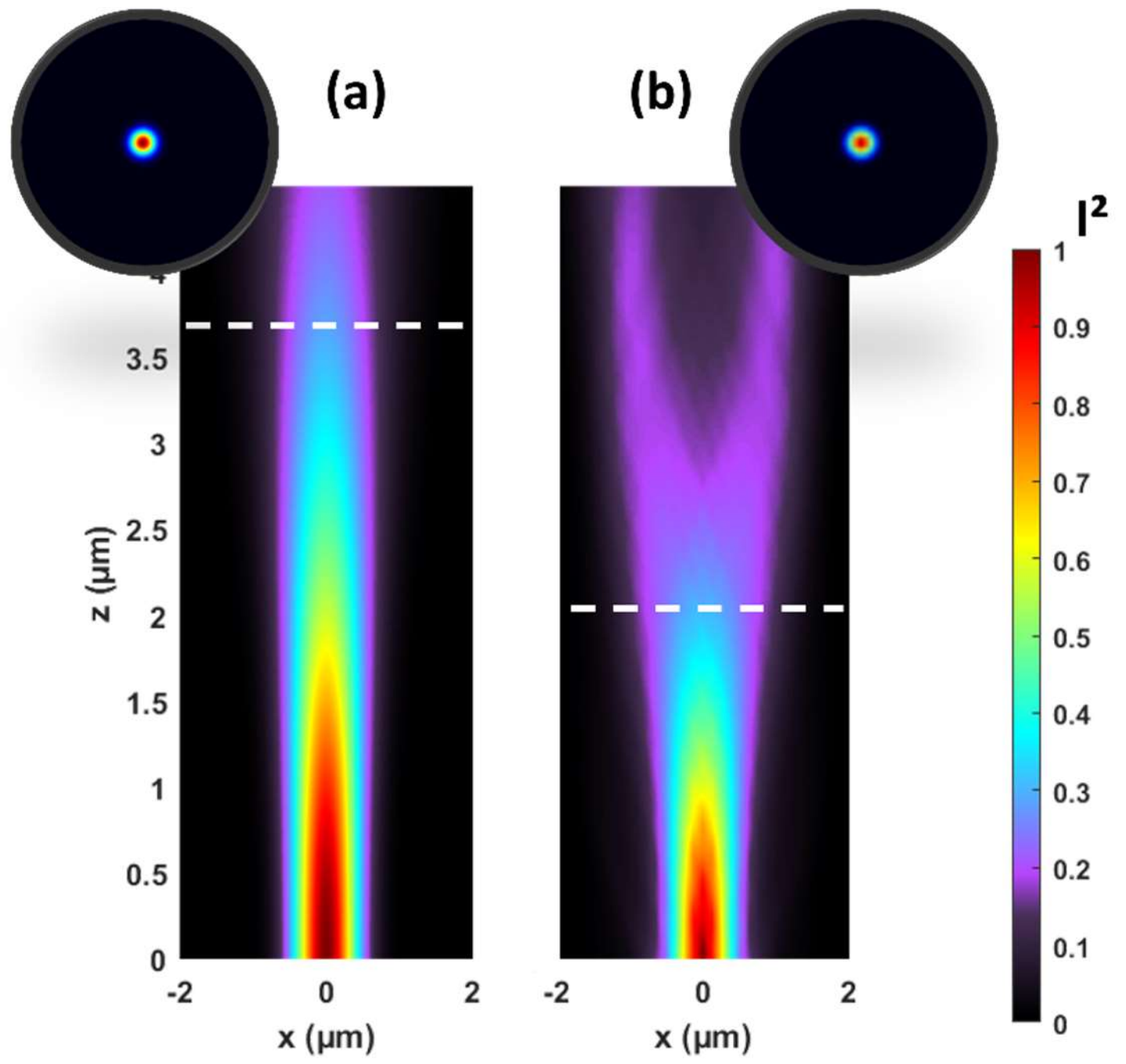

Fig. 2. Squared intensity from the propagation of Gaussian (a) or multimode (b) beam in free space; Dashed lines represent the limit of a power density that permits efficient two-photon fluorescence generation. The round insets show the squared intensity of Gaussian (a) and multimode (b) beams in the far field.

## 3. Experimental setup

The setup that we used to perform our two-photon imaging experiments consists of a femtosecond laser pump (Light Conversion Pharos) emitting infrared radiation at 1030 nm, with a temporal pulse duration tunable between a minimum of 174 fs up to a few tens of picoseconds. The laser frequency repetition rate can also be varied from 1 kHz to 200 kHz. We used a custom-made (Prospective Instruments) multiphoton microscope with no laser system incorporated inside. This permitted us to properly shape the laser beam for our experiments. We performed all characterizations with the shortest available femtosecond pulses, to minimize the thermal damages on the samples (fluorescent polystyrene microbeads-Invitrogen TetraSpeck) [21], with a repetition frequency equal to or higher than 100 kHz. To reshape the laser Gaussian beam, we used a commercial Alcatel GRIN MMF fiber, with a core/cladding diameter of 50/125 μm. The necessary fiber length was 5 cm, in order to generate a proper multimode beam, composed of a sufficient number of radial modes. The laser beam was focused inside the fiber through a 50 mm converging lens. At the fiber output, the beam was collimated through an 11 mm converging lens.

## 4. Experimental results

### *4.1 Rayleigh length comparison between the Gaussian and multimode beams*

In the first step, we compared the divergence in free space of Gaussian and multimode beams. For doing that, we focalized the beams through a 50 mm converging lens, and used a CCD camera on a translation stage, so that we could analyze the spatial beam profile upon distance along the propagation axis (see Fig. 3). The multimode beam size, at the output face of the 50/125 µm GRIN fiber, is of 10 µm (FWHMI) approximately. After passing through the 11 mm/50 mm lens system, the beam is magnified by a factor of 4.5, resulting in a size of approximately 45 µm at the focal point (FWHMI). The Gaussian beam was obtained without MMF fiber, directly from the laser, and adjusted to match the size of the multimode beam (FWHMI) on the camera, using an additional afocal system (50 mm / 11 mm converging lenses duet).

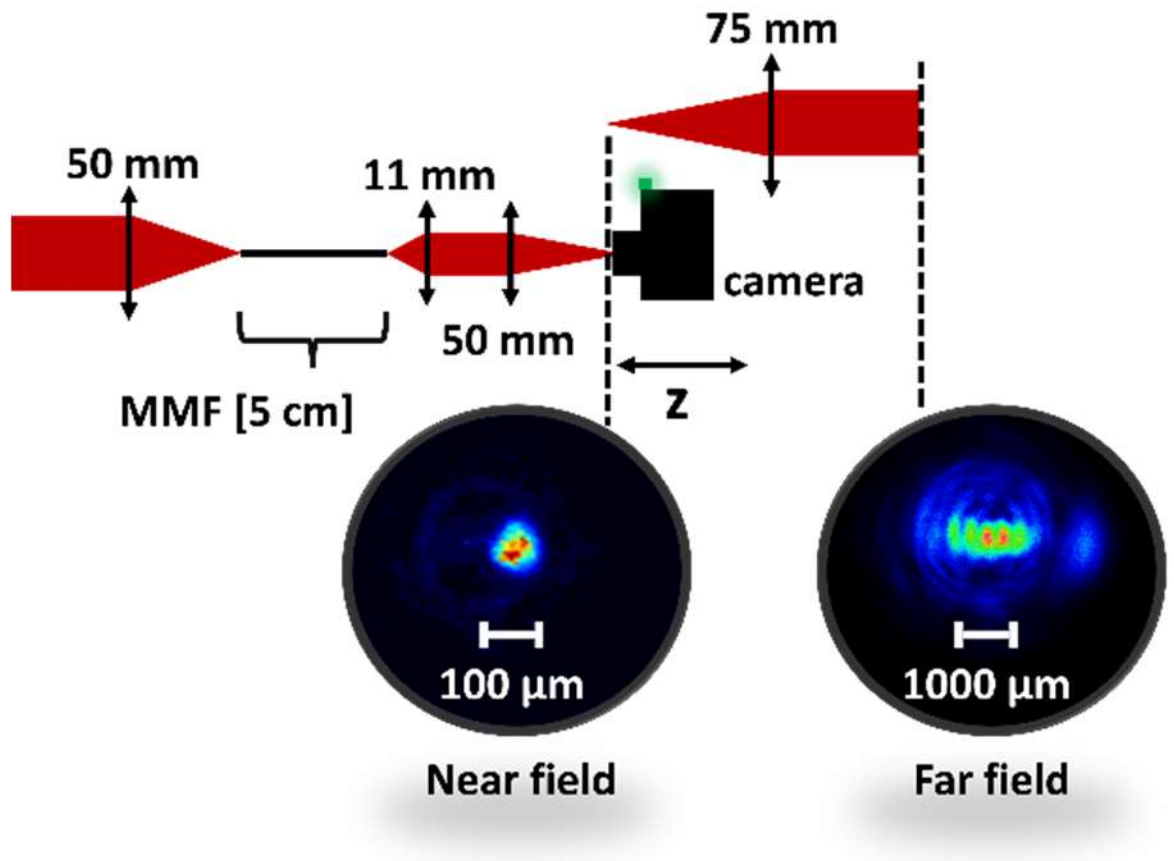

Fig. 3. Experimental setup for recording near and far-fields, after beam propagation in the GRIN multimode fiber. Images of the shaped multimode beam (near and far field) are shown in the round insets.

In order to experimentally prove that the multimode-shaped beam has a Rayleigh length shorter than that of the reference Gaussian beam, we scanned the beam transversal shape at different points in space along the propagation axis, by using a CCD camera and a translation stage (the setup is visible on Fig. 3). As we can see on the round insets of Fig. 3, the multimode beam profile looks almost Gaussian in the near-field, meaning that it will still be Gaussian when focalized through the microscope objective on the sample, ensuring a good transverse resolution, as in two-photon microscopy when using Gaussian laser beams. In Fig. 4, we plotted the evolution of the diameter (FWHMI) of both the Gaussian and the multimode beams, versus propagation length z, by using the setup of Fig. 3. The diameter of the multimode beam (Fig. 4 (b)) increases much more than that of the Gaussian beam (Fig. 4 (a)) when z grows larger (z = 0 is set at the minimum beam waist). This is a proof that the divergence of the multimode beam is higher (hence the Rayleigh range is shorter) than that of the Gaussian beam, when both beams propagate in free space, and this for similar beam diameters at the focal point (z = 0). Indeed, we found a Rayleigh range of 4 mm for the Gaussian beam, versus 1 mm for the multimode beam. The calculated $M^2$ is 2 for the radial multimode beam (versus 1 for a Gaussian beam). The decrease in Rayleigh length that we observed experimentally is larger than in our previous simulation (see Fig. 2). This can be explained by the higher number of modes constituting the multimode beam that we experimentally obtained, when compared with the simulated multimode beam. The shape of the beam is also affected by this modal composition difference, as we can see by comparing the rounds insets of Fig. 1 and Fig. 3. The beam that we obtained experimentally has a larger asymmetry in the far-field, which can be caused by the excitation of a few radially asymmetric modes in the fiber. Despite this asymmetry, the shape of our generated multimode beam is still Gaussian in the near-field, so that it can be effectively exploited for two-photon microscopy imaging.

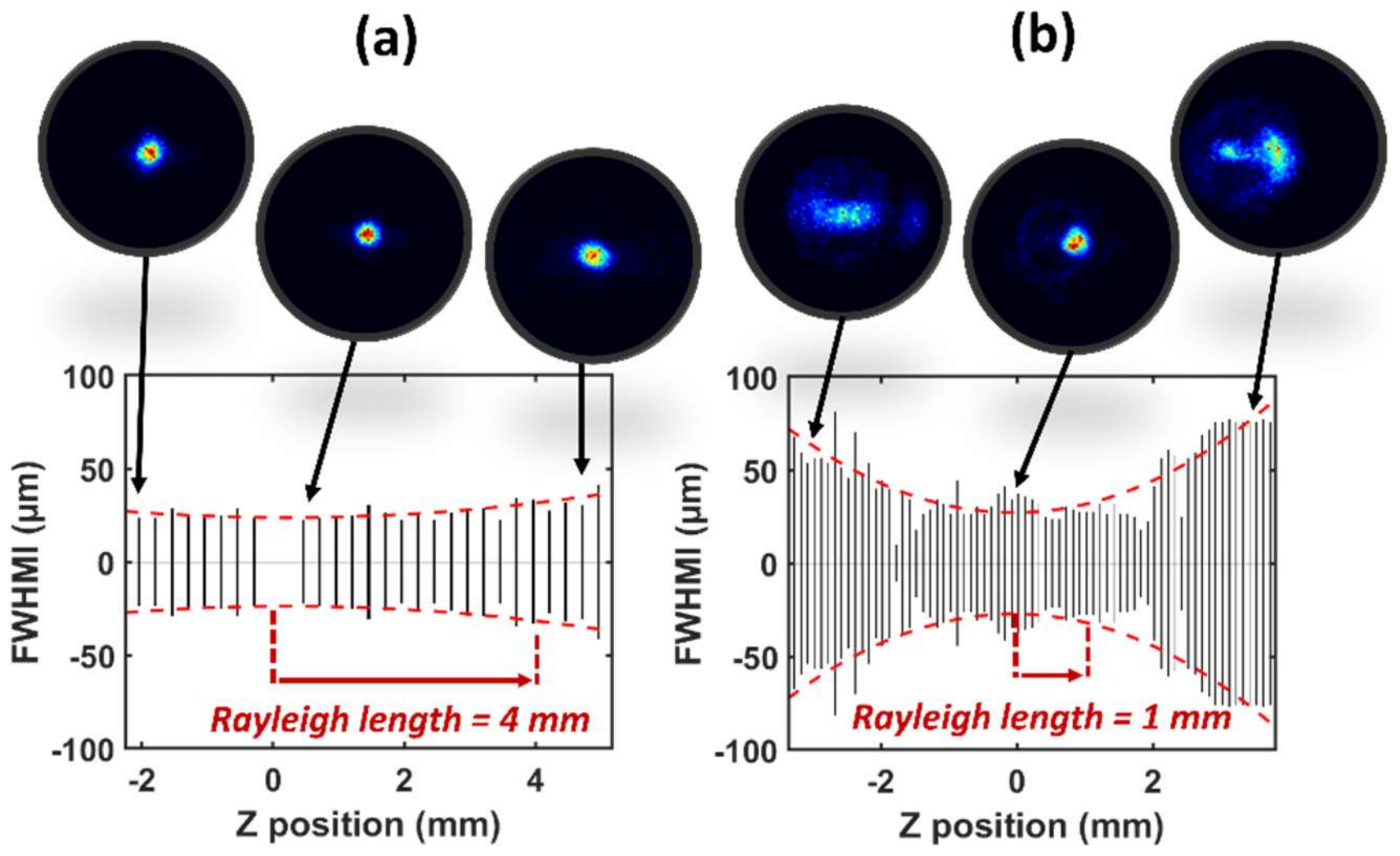

Fig. 4. FWHMI evolution for (a) the Gaussian beam, and (b) the radial multimode beam. These plots are accompanied by images of the beams for different values of z (round insets), to show the evolution of the beam shape along the propagation distance.

### *4.2 Axial resolution improvement of the multiphoton microscope*

Next, we injected the beam generated through the GRIN MMF into the multiphoton microscope. An afocal system, composed of a 200 mm and a 75 mm converging lens, was used to downsize the beam to 300 µm of diameter (FWHMI), before its injection in the microscope, so that we could enhance the fluorescence signal from the sample. The laser beam was focalized on the sample through a 20x microscope objective (with a large numerical aperture of 1 in water).

The laser pulse sent into the sample was a 174 fs pulse with an energy of 6 nJ. This energy level is sufficient for conducting two-photon fluorescence imaging of highly fluorescent polystyrene beads. It is important to keep in mind that the signal from a different sample can be enhanced by increasing the pulse energy further. Notably, the only limitation to increasing the pulse energy is the presence of silica nonlinearities, but their impact remains low when using 5 cm of fiber. We observed that the damage threshold for 1 µm polystyrene beads was approximately 10 nJ, which is significantly higher than our current pulse energy range.

The fluorescent signal coming from each point of the sample was collected in epi configuration, and measured by photon counters. A numerical image reconstruction was then implemented, in order to obtain a 3D image of the sample. We injected first the shaped radial multimode beam, and then the Gaussian beam in the multiphoton microscope. To measure the longitudinal resolution improvement brought about by the shaping of the multimode laser beam, we measured the point spread function (PSF) for both cases, i.e., with either a shaped multimode beam or a Gaussian beam. This PSF was obtained by stacking 3D images of 1 µm diameter fluorescent polystyrene beads (see Fig. 5). Several sets of experiments and images showed that better resolved 3D images were obtained by replacing the initial converging lens (50 mm) that was used to focalize the laser inside the short piece of fiber, with a 25 mm converging lens, in order to increase the numerical aperture of the laser beam that is injected in the fiber, thus exciting more fiber modes [22]. Before the fiber, we also added a half-wave plate for the control of the beam polarization. Thus, we managed to obtain the most symmetrical multimode beam of our experiments, that is visible on the round insets of Fig. 5 (b). With this newly shaped beam, we could observe an almost twofold shrinking of the polystyrene bead signature along the z axis (Fig. 5 (b)), when compared with the case of a Gaussian beam (Fig. 5 (a)). This spatial compression results from the neat improvement of the axial resolution of our microscopy system, decreasing from 13 µm with a Gaussian beam (Fig. 5 (c)), to only 7 µm when using the shaped multimode beam (Fig. 5 (d)) that we show in the round inset of Fig. 5 (b). The transversal resolution, along the x and y axes, remains comparable for both beams. The drawback of using a multimode beam is that the signal from the sample is lower than that from a Gaussian beam. This deterioration in the signal is clearly due to the improvement in longitudinal resolution, i.e. the shortening of the interaction of the beam with the sample. Moreover, the fact that the multimode beam is not perfectly Gaussian in the near field can also introduce additional energy scattering at the beam periphery. Interferences between the different modes of the multimode beam might create a fringe pattern along the axial direction, resulting in axial side lobes on the 3D image. However, no side lobes were observed in our imaging experiments. In our experiments, all the beads are similar, taken from the same sample plate, but images are from different spots of the plate, because of the need to realign the microscope when modifying the beam.

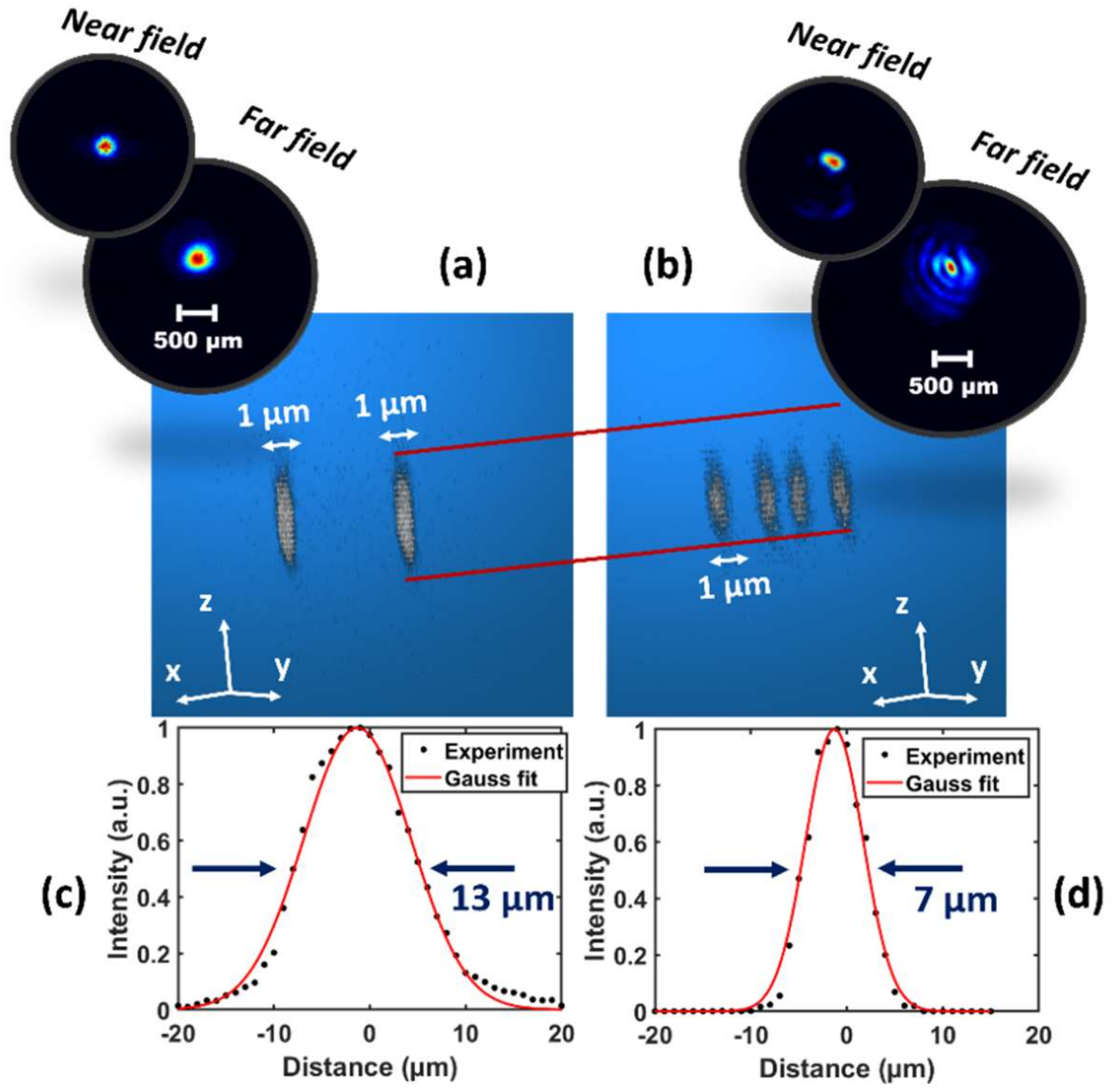

Fig. 5. Top: 3D images of 1 µm polystyrene beads obtained with a Gaussian beam (a) or with a multimode beam (b). Bottom: axial resolution of our microscope, when using the Gaussian beam (c) or the multimode beam (d). The near and far-field images of the beams are visible on the round insets.

## 5. Conclusions

We demonstrate that it is possible to generate a tailored multimode beam in the linear propagation regime, mostly composed of a sum of radial modes, by means of a short piece of MMF GRIN fiber, in which we injected a Gaussian beam. This simple and low-cost method allowed us to generate radially symmetric multimode beams which, when used as a light source for a multiphoton microscope, could achieve a twofold improvement of the axial resolution of 3D images, without any degradation of their transversal resolution. Our method also holds potential for applications in two-photon polymerization, which would benefit of the enhanced axial resolution.

**Funding.** Nationale de la Recherche (ANR-10-LABX-0074-01), Italian Ministry of Health: RF-2021-12373094, European Union - NextGenerationEU through the Italian Ministry of University and Research under PNRR - M4C2-I1.3 Project PE_00000019 "HEAL ITALIA" CUP B53C22004000006, Sapienza Università di Roma AR12218166F0555A.

**Acknowledgments.** R. Jauberteau thanks M. Jonard for his advice on microscopy and imaging technics.

**Disclosures.** The authors declare no conflicts of interest.

**Data availability.** Data underlying the results presented in this paper are not publicly available at this time but may be obtained from the authors upon reasonable request.

## References

1. A. Picozzi, G. Millot, and S. Wabnitz, "Nonlinear virtues of multimode fibre," Nature Photon **9**(5), 289-291 (2015).
2. K. Krupa, A. Tonello, A. Barthélémy, *et al.*, "Multimode nonlinear fiber optics, a spatiotemporal avenue," APL Photonics **4**(11), 110901 (2019).
3. E. Deliancourt, M. Fabert, A. Tonello, *et al.*, "Wavefront shaping for optimized many-mode Kerr beam self-cleaning in graded-index multimode fiber," Opt. Express **27**(12), 17311-17321 (2019).
4. G. Lopez-Galmiche, Z. Sanjabi Eznaveh, M. A. Eftekhar, *et al.*, "Visible supercontinuum generation in a graded index multimode fiber pumped at 1064 nm," Opt. Lett. **41**(11), 2553-2556 (2016).
5. S. Norman, M. Zervas, A. Appleyard, *et al.*, "Power scaling of high-power fiber lasers for micromachining and materials processing applications," Fiber Lasers III: Technology, Systems, and Applications **6102**, 402-408 (2006).
6. G. Li, N. Bai, N. Zhao, *et al.*, "Space-division multiplexing: the next frontier in optical communication," Adv. Opt. Photon. **6**(4), 413-487 (2014).
7. N. O. Moussa, T. Mansuryan, C.-H. Hage, *et al.*, "Spatiotemporal beam self-cleaning for high-resolution nonlinear fluorescence imaging with multimode fiber," Sci Rep **11**(1), 18240 (2021).
8. Y. Zhang, Y. Cao, and J.-X. Cheng, "High-resolution photoacoustic endoscope through beam self-cleaning in a graded index fiber," Opt. Lett. **44**(15), 3841-3844 (2019).
9. O. Tzang, A. M. Caravaca-Aguirre, K. Wagner, *et al.*, "Adaptive wavefront shaping for controlling nonlinear multimode interactions in optical fibres," Nature Photon **12**(6), 368-374 (2018).
10. I. M. Vellekoop and A. P. Mosk, "Focusing coherent light through opaque strongly scattering media," Opt. Lett. **32**(16), 2309-2311 (2007).
11. T. Čižmár, M. Mazilu, and K. Dholakia, "In situ wavefront correction and its application to micromanipulation," Nature Photon **4**(6), 388-394 (2010).
12. K. Krupa, A. Tonello, B. M. Shalaby, *et al.*, "Spatial beam self-cleaning in multimode fibres," Nature Photon **11**(4), 237-241 (2017).
13. E. Deliancourt, M. Fabert, A. Tonello, *et al.*, "Kerr beam self-cleaning on the LP 11 mode in graded-index multimode fibers," OSA Continuum **2**(4), 1089-1096 (2019).
14. M. Göppert - Mayer, "Über elementarakte mit zwei quantensprüngen," Annalen der Physik **401**(3), 273-294 (1931).
15. W. Kaiser, and CGB. Garrett, "Two-photon excitation in Ca F 2: Eu 2+," Physical Review Letters **7**(6), 229 (1961).
16. J. Tang, J. Ren, and K. Y. Han, "Fluorescence imaging with tailored light," Nanophotonics **8**(12), 2111-2128 (2019).
17. S. N. Khonina, N. L. Kazanskiy, S. V. Karpeev, *et al.*, "Bessel beam: Significance and applications—A progressive review," Micromachines **11**(11), 997 (2020).
18. T. A. Planchon, L. Gao, D. E. Milkie, *et al.*, "Rapid three-dimensional isotropic imaging of living cells using Bessel beam plane illumination," Nat Methods **8**(5), 417-423 (2011).
19. X. Zhu, A. Schülzgen, H. Li, *et al.*, "Coherent beam transformations using multimode waveguides," Opt. Express **18**(7), 7506-7520 (2010).
20. E. Guevara "Finite Difference Beam Propagation Method," (https://www.mathworks.com/matlabcentral/fileexchange/14887-finite-difference-beam-propagation-method), MATLAB Central File Exchange (2023).
21. E. E. Hoover and J. A. Squier, "Advances in multiphoton microscopy technology," Nature Photon **7**(2), 93-101 (2013).
22. R. Paschotta, "Beam quality limit for multimode fibers," https://www.rp-photonics.com/spotlight_2013_11_12.html .